\documentstyle [12pt,epsf,rotate] {article}
\newcommand{\nwc}{\newcommand}
\nwc{\be}  {\begin{equation}}
\nwc{\ee}  {\end{equation}}
\nwc{\ba}  {\begin{array}}
\nwc{\ea}  {\end{array}}
\nwc{\bdm} {\begin{displaymath}}
\nwc{\edm} {\end{displaymath}}
\nwc{\bea} {\be\ba{rcl}}
\nwc{\eea} {\ea\ee}
\nwc{\bear} {\begin{eqnarray}}
\nwc{\ear} {\end{eqnarray}}

\begin{document}
\begin{titlepage}
\begin{flushright}
HD--THEP--99--34
\end{flushright}
\quad\\
\vspace{1.8cm}

\begin{center}
{\Large Gluon-Meson Duality in the Mean Field Approximation}\\
\vspace{2cm}
Christof Wetterich\footnote{e-mail: C.Wetterich@thphys.uni-heidelberg.de}\\
\bigskip
Institut  f\"ur Theoretische Physik\\
Universit\"at Heidelberg\\
Philosophenweg 16, D-69120 Heidelberg\\
\vspace{3cm}

\end{center}

\begin{abstract}
In a gauge-fixed language gluon-meson duality can be described
as the Higgs mechanism for ``spontaneous symmetry breaking''
of color. We present a mean field computation which suggests that
this phenomenon is plausible in QCD. One obtains realistic masses
of the light mesons and baryons.
\end{abstract}
\end{titlepage}

A dual description of QCD in the low momentum or strong coupling
region should contain the light meson and baryon degrees of freedom
as the relevant fields. This is opposed to the quark-gluon description
in the high momentum or weak coupling regime. We have proposed recently
\cite{GM}, \cite{RU} that for three light flavors such a dual description indeed
exists, with the light gluon fields associated to the light vector
meson octet and the nine quark fields to the light baryon octet and
a singlet. (For some related ideas\footnote{In contrast to \cite{E},
we consider here the ground state of standard QCD, without
additional ``fundamental'' light-colored scalar fields
and without additional gauge interactions. We discuss the
vacuum and do not deal with the properties of QCD at very high baryon
density.} see \cite{E}, \cite{SW}.) This picture realizes gluon-meson
duality and quark-baryon duality in a straightforward
way. Already a very simple form of an effective action can
account for realistic masses of the light baryons and the light
pseudoscalar and vector meson octets. It also leads to
realistic interactions of the Goldstone bosons and
predicts the couplings of the vector mesons to pions and
baryons in the observed range.

The key ingredient of this picture is a condensate of a suitable
quark-antiquark operator which is associated to a color octet.
Whereas in \cite{GM} we have mainly used a gauge-invariant language
in terms of nonlinear fields we concentrate in this note on
a (linear) gauge-fixed version (preferably in Landau gauge).
In this version the $SU(3)$-color group is ``spontaneously
broken'' by the expectation value of the color octet
condensate\footnote{We use the language of spontaneous
symmetry breaking similar to the electroweak theory, despite the
fact that gauge symmetries are never broken in a strict sense.}.
The relevant quark-antiquark condensate also transforms as an
octet under the vectorlike $SU(3)$ flavor group. Its expectation
value conserves a diagonal global $SU(3)$ symmetry of combined
color and flavor rotations. This is associated with the physical
flavor group of the ``eightfold way''. Due to this ``color-flavor
locking'' \cite{SW} the quark fields transform as an octet and a singlet
under the physical flavor group. They carry the appropriate integer
electric charges and can be associated with baryons\footnote{The
gauge fixing is not compatible with the phase transformations
corresponding to baryon number. In the gauge-invariant language
with nonlinear fields it can be verified that baryons carry
indeed three times the baryon number of the quarks \cite{RU}.}. The octet of
gluons acquires a mass through the Higgs mechanism and can be
identified with the light vector mesons\footnote{We note, however,
that the $\phi$-meson and the appropriate mixing of vector mesons
is not yet contained in the simplest version. See \cite{RU} for an extension  
with addition of the singlet vector meson.}. Chiral symmetry
is spontaneously broken by the octet condensate, as well as by
the usual singlet condensate. In the absence of quark masses this
leads to eight Goldstone bosons.

In \cite{GM} we have mainly discussed the consequences of an
assumed expectation value of a color octet quark-antiquark bilinear.
Here we present a first investigation if it is plausible that such a
vacuum expectation value is generated dynamically in QCD. In this first
step we want to identify the mechanisms which could lead to dynamical
spontaneous color symmetry breaking. We do not yet intend
quantitative estimates.

For this purpose we treat the quantum fluctuations of the light
baryons or quarks and the light mesons in the mean field approximation.
We consider the non-perturbative region of momenta $q^2<k^2$,
with $k$ an appropriate cutoff (typically $k=850$ MeV).
The form of the effective one-particle irreducible multiquark
interactions at the scale $k$ is largely dictated by chiral
symmetry, color symmetry and the discrete symmetries $P$ and $C$,
as well as by the known form of the axial anomaly. We include
effective interactions involving up to eight quarks or antiquarks.
The coefficients of the corresponding invariants
in the effective action at the scale
$k$ are treated, however, as free parameters. We demonstrate
that a reasonable choice of these couplings leads indeed to
spontaneous color symmetry breaking and realistic
masses for all light particles. This is largely
due to the fermion fluctuations with momenta
$0<q^2<k^2$. They are cut off by mass terms in case
of chiral symmetry breaking, whereas their contribution
disfavors a ground state with absence of octet and singlet condensates.
A second crucial ingredient are cubic couplings which
reflect the chiral $U_A(1)$ anomaly. A future QCD calculation will
have to determine the values of the multiquark couplings at
the scale $k$. Only after this second step one may draw a definite
conclusion about the realization of spontaneous color symmetry
breaking in QCD.

It is convenient to introduce for the color singlet and octet
quark-antiquark bilinears the notation
\bear\label{1}
\tilde\varphi^{(1)}_{ab}&=&\bar\psi_{L\ ib}\ \psi_{R\ ai}\quad,\quad
\tilde\varphi^{(2)}_{ab}=-\bar\psi_{R\ ib}\ \psi_{L\ ai}
\nonumber\\
\tilde\chi^{(1)}_{ij,ab}&=&\bar\psi_{L\ jb}\ \psi_{R\ ai}
-\frac{1}{3}\bar\psi_{L\ kb}\ \psi_{R\ ak}\ \delta_{ij}\nonumber\\
\tilde\chi^{(2)}_{ij,ab}&=&-\bar\psi_{R\ jb}\ \psi_{L\ ai}
+\frac{1}{3}\bar\psi_{R\ kb}\ \psi_{L\ ak}\ \delta_{ij}\ear
where $i,j, k$ are color indices and $a,b$ refer to flavor. With
respect to the chiral flavor group $SU(3)_L\times SU(3)_R$,
the bilinears $\tilde\varphi^{(1)}$ and $\tilde\chi^{(1)}$
transform as $(\bar 3,3)$ whereas $\tilde\varphi^{(2)}$ and $\tilde
\chi^{(2)}$ are in the $(3,\bar 3)$ representation. Parity maps $\tilde
\varphi^{(1)}\leftrightarrow\tilde\varphi^{(2)}$, $\tilde\chi^{(1)}
\leftrightarrow\tilde\chi^{(2)}$ whereas under charge conjugation the
transformation is $\tilde\varphi^{(i)}\leftrightarrow\tilde\varphi^{(i)T}$,
$\tilde\chi^{(i)}_{ij,ab}\leftrightarrow\tilde\chi^{(i)}_{ji,ba}$.
At the cutoff
we consider an effective Lagrangian of the form\footnote{Gluons
are incorporated by a covariant derivative of the quarks and an
appropriate kinetic term involving their field strength.}
\be\label{2}
{\cal L}_k=i\bar\psi_{ia}\gamma^\mu\partial_\mu\psi_{ai}-
\tilde U_k(\tilde\varphi,\tilde\chi)+{\cal L}_{MK}+{\cal L}_\eta\ee
with
\bear\label{3}
&&\tilde U_k(\tilde\varphi,\tilde\chi)=2\lambda_\sigma\tilde\rho
+2\lambda_\chi\tilde\rho_\chi+\tau_\sigma\tilde\rho^2+\tau_\chi\tilde
\rho_\chi^2+\tau_\gamma\tilde\rho\tilde\rho_\chi\nonumber\\
&&+\zeta\left\{det\tilde\varphi^{(1)}
+det\tilde\varphi^{(2)}+
E(\tilde\varphi^{(1)},\tilde\chi^{(1)})+E(\tilde\varphi^{(2)},
\tilde\chi^{(2)})\right\}\ear
Here the multiquark interactions $\tilde U_k$ are expressed in terms of
the chirally invariant color singlets
\be\label{4}
\tilde\rho=\tilde\varphi^{(1)}_{ab}\ \tilde\varphi^{(2)}_{ba}\quad,
\quad
\tilde\rho_\chi=\tilde\chi^{(1)}_{ij,ab}\ \tilde\chi_{ji,ba}^{(2)}\ee
and the 't Hooft term for the chiral anomaly \cite{H} (with
coefficient $\zeta$) with
\be\label{5}
E(\tilde\varphi,\tilde\chi)=\frac{1}{6}\epsilon_{a_1a_2a_3}
\epsilon_{b_1b_2b_3}\tilde\varphi_{a_1b_1}\tilde\chi_{ij,a_2b_2}
\tilde\chi_{ji,a_3b_3}\ee
We assume that the quantum fluctuations with momenta $q^2>k^2$ have
already been integrated out, such that the remaining functional
integral has an effective ultraviolet cutoff $k$.

In QCD perturbation theory the one-particle-irreducible four-quark
interactions $\sim\tilde\rho,\tilde\rho_\chi$ are generated
by box diagrams with
\be\label{6}
\lambda_{\sigma,\chi}=\frac{L_{\sigma,\chi}}{32\pi^2}
\frac{g^4}{k^2}\quad,\quad L_\sigma=\frac{23}{9}l^4_3\quad,
\quad L_\chi=\frac{13}{24}l^4_3\ee
Here $k$ is the infrared cutoff for the loop momenta and the constant
$l^4_3$ is about one, its value depending on the precise
implementation of the cutoff. The coefficients $\tau_l$ would
correspond to eight quark interactions generated by diagrams
with four gluons, $\tau_l\sim g^8/(32\pi^2k^8)$.
We will use, however, a scale $k\approx 850$ MeV in the
non-perturbative domain and treat for the present work the couplings
$\lambda_l$ and $\tau_l$ as free parameters. For the 't Hooft
interaction we have used an appropriate Fierz transformation,
and $\zeta$ is again treated as a free parameter. The terms
${\cal L}_\eta$ and ${\cal L}_{MK}$ contain sources  for the
quarks and quark-antiquark bilinears
\bear\label{7}
{\cal L}_\eta&=&-\bar\eta\psi-\eta\bar\psi\nonumber\\
{\cal L}_{MK}&=&-M_{ba}\tilde\varphi_{ab}^{(2)}-M^\dagger_{ba}\tilde\varphi
^{(1)}_{ab}-K_{ij,ab}\tilde\chi^{(2)}_{ji,ba}-K^*_{ij,ab}
\tilde\chi^{(1)}_{ij,ab}\ear
The physical situation corresponds to $\eta=\bar\eta=0$,
$K_{ij,ab}=0$, $M_{ab}(x)=diag(m_u,m_d,m_s)$
with $m_q$ the (current) quark masses.

The first step in a mean field discussion of the
multi-quark action, $S=\int d^4x{\cal L}_k$,
is partial bosonization. We introduce a factor of unity (up to an
irrelevant overall constant) in the functional integral for the
partition function
\bear\label{8}
&&Z[M,K,\eta,\bar\eta]=\int D\psi D{\bar\psi}\exp(-S)\\
&=&\int D\psi D\bar\psi D\sigma D\xi\exp\{-S\nonumber\\
&&-\int d^4x[U_k(\sigma^R
_{ab}-\tilde\varphi^R_{ab},\sigma^I_{ab}-\tilde\varphi^I_{ab},
\xi^R_{ij,ab}-\tilde\chi^R_{ij,ab},\xi^I_{ij,ab}-
\tilde\chi^I_{ij,ab})\nonumber\\
&&+{\cal L}_{MK}(\sigma^R_{ab}-\tilde\varphi_{ab}^R,\sigma^I_{ab}
-\tilde\varphi^I_{ab},\xi^R_{ij,ab}-\tilde\chi^R_{ij,ab}
,\xi^I_{ij,ab}-\tilde\chi^I_{ij,ab})]\}\nonumber\ear
with
\bear\label{9}
\tilde\varphi^R_{ab}&=&\frac{1}{2}(\tilde\varphi_{ab}^{(1)}+\tilde
\varphi_{ba}^{(2)})\quad,\quad\quad\hfill\tilde\varphi_{ab}^I\
=-\frac{i}{2}(\tilde\varphi^{(1)}_{ab}-\tilde\varphi^{(2)}_{ba})
\nonumber\\
\tilde\chi^R_{ij,ab}&=&\frac{1}{2}(\tilde\chi^{(1)}_{ij,ab}+\tilde
\chi^{(2)}_{ji,ba})\quad,\quad\hfill\tilde\chi^I_{ij,ab}=-\frac{i}{2}
(\tilde\chi^{(1)}_{ij,ab}-\tilde\chi^{(2)}_{ji,ba})\ear
The function $U_k$ is determined by the requirement that
$U_k(-\tilde\varphi^R,-\tilde\varphi^I,-\tilde\chi^R$,$-\tilde\chi^I)
=\tilde U_k(
\tilde\varphi^R,\tilde\varphi^I,\tilde\chi^R,\tilde\chi^I)$
is given by (\ref{3})
expressed in the appropriate variables.

The effective Lagrangian of the bosonized theory
\be\label{10}
{\cal L}^{(B)}={\cal L}(\tilde\varphi,\tilde\chi)+U_k(\sigma-\tilde\varphi
,\xi-\tilde\chi)+{\cal L}_{MK}(\sigma-\tilde\varphi,\xi-\tilde\chi)\ee
is next expanded in powers of $\tilde\varphi$ and $\tilde\chi$.
The terms involving no fermion fields result in masses, sources, and
interactions for the scalars
\be\label{11}
{\cal L}_s=U_k(\sigma,\xi)-Tr(M^\dagger\sigma+M\sigma^\dagger)
-(K^*_{ij,ab}\xi_{ij,ab}+K_{ij,ab}\xi^*_{ij,ab})\ee
The potential $U_k$ obtains from (\ref{3}) by the replacements
$\tilde\varphi^{(1)}_{ab}\to\sigma_{ab},\
\tilde\varphi^{(2)}_{ab}\to\sigma^\dagger_{ab},\ \tilde\chi
^{(1)}_{ij,ab}\to\xi_{ij,ab},\ \tilde\chi^{(2)}_{ij,ab}\to\xi^*_{ji,ba}$,
$\tilde\rho\to Tr\ \sigma^\dagger\sigma,\  
\tilde\rho_\chi\to\xi^*_{ij,ab}\xi_{ij,ab}$, where
we have combined $\sigma_{ab}=\sigma_{ab}^R+i\sigma^I_{ab},
\ \xi_{ij,ab}=\xi^R_{ij,ab}+i\xi^I_{ij,ab}$. Furthermore, the
sign of the term $\zeta$ is switched. The chiral  and discrete
transformation properties carry over from $\tilde\varphi,\tilde\chi$
to $\sigma,\xi$ and $U_k$ is therefore invariant under the
corresponding symmetries. The inclusion of the eight-quark interactions
in (\ref{3}) guarantees that for positive $\tau_l$ the
functional integral (\ref{8}) is well defined and $U_k$ is bounded
from below. On the other hand, the terms in the expansion of
$U_k+{\cal L}_{MK}$ which
involve only fermion fields precisely cancel the multiquark
interactions $-\tilde U_k$ and the source ${\cal L}_{MK}$ for the
quark bilinears. Chiral symmetry breaking due to quark masses appears
now in the form of a linear source term for the scalar fields
(\ref{11}). The terms linear in the quark bilinears give rise to
Yukawa-type interactions, involving a quark-antiquark pair and one or
several scalar fields. Finally, the terms $\sim\zeta$
and $\tau_l$ also produce interactions between four or six
fermions and one or two
scalars. In our mean field approximation these residual
interactions involving more than two quark or antiquark fields are
neglected.

We want to perform the remaining fermionic functional integral
for a scalar background field which preserves a ``diagonal''
vector-like
$SU(3)$ symmetry and is invariant under $C$ and $P$
\be\label{12}
\sigma_{ab}=\bar\sigma\delta_{ab}\quad,\quad\xi_{ij,ab}
=\frac{1}{\sqrt6}\bar\xi(\delta_{ia}\delta_{jb}-\frac{1}{3}
\delta_{ij}\delta_{ab})\ee
For this configuration\footnote{We take both $\bar\sigma$ and
$\bar\xi$ to be real such that $P$ is conserved. In principle,
$\bar\xi$ could have a relative phase as compared to $\bar\sigma$.
A purely imaginary $\bar\xi$ is favored by the anomaly term in
$U_k$ if $\zeta\bar\sigma$ is positive at the minimum. From this
point of view one may prefer a negative value of $\zeta$. On the
other hand, invariants of the type $\tilde\rho\tilde\rho_\chi-
\frac{1}{4}(\sigma^\dagger_{ab}\xi_{ijbc}\sigma^\dagger_{cd}
\xi_{jida}+c.c.)$ can also favor real $\bar\xi$. We have not
included in the present discussion invariants which vanish identically
for the configuration (\ref{12}) and real $\bar\sigma,\bar\xi$ as the
one above. Parity conservation remains an issue in our approach. In order
to settle this issue one needs to include invariants which contribute
to a background field (\ref{12}) with an arbitrary phase for
$\bar\xi$. Also the quantum fluctuations will have to be treated
in this more general setting. At the present stage the positive mass term
$M^2_{\eta'}$ for the $\eta'$-meson (see below) is a strong
hint that the ground state preserves parity.}
the effective Lagrangian (\ref{10}) is
given by
\be\label{13}
{\cal L}^{(B)}=U_{cl}(\bar\sigma,\bar\xi)+i\bar\psi_{ia}
\gamma^\mu\partial_\mu\psi_{ai}
+M_8(\bar\sigma,\bar\xi)\bar\psi_{ia}^{(8)}\bar\gamma\psi_{ai}
^{(8)}+M_1(\bar\sigma,\bar\xi)\bar\psi^{(1)}\bar\gamma\psi^{(1)}
\ee
with $\bar\gamma$ the Euclidean analogue\footnote{ For
our conventions see \cite{CW}.} of $\gamma^5$.
The fermionic part involves a mass term for the $SU(3)$-singlets
and octets
\be\label{14}
\psi^{(1)}=\frac{1}{\sqrt3}\psi_{aa}\quad,\quad \psi^{(8)}_{ai}
=\psi_{ai}-\frac{1}{\sqrt3}\psi^{(1)}\delta_{ai}\ee
where the latter is identified with the light baryon octet.
One finds
\bear\label{15}
M_8&=&2\lambda_\sigma\bar\sigma-\frac{2}{3\sqrt6}\lambda_\chi\bar\xi-
\zeta(\bar\sigma^2-\frac{2}{27}\bar\xi^2+\frac{1}{
9\sqrt6}\bar\sigma\bar\xi)\nonumber\\
&&+6\tau_\sigma\bar\sigma^3+\frac{4}{3}\tau_\gamma\bar\xi^2\bar
\sigma-\frac{1}{\sqrt6}\tau_\gamma\bar\sigma^2\bar\xi
-\frac{8}{9\sqrt6}\tau_\chi
\bar\xi^3\ear
\bear\label{16}
M_1&=&2\lambda_\sigma\bar\sigma+\frac{16}{3\sqrt6}\lambda_\chi
\bar\xi-\zeta\left(
\bar\sigma^2-\frac{2}{27}\bar\xi^2-\frac{8}{9\sqrt6}
\bar\sigma\bar\xi\right)
\nonumber\\
&&+6\tau_\sigma\bar\sigma^3+\frac{4}{3}\tau_\gamma
\bar\xi^2\bar\sigma+\frac
{8}{\sqrt6}\tau_\gamma\bar\sigma^2\bar\xi+\frac{64}{9\sqrt6}
\tau_\chi\bar\xi^3\ear
The classical scalar potential reads
\bear\label{17}
U_{cl}&=&-2m\bar\sigma+6\lambda_\sigma\bar\sigma^2+\frac{8}{3}
\lambda_\chi\bar\xi^2
-2\zeta\bar\sigma^3+\frac{4\zeta}
{9}\bar\sigma\bar\xi^2\nonumber\\
&&+9\tau_\sigma\bar\sigma^4+\frac{16}{9}\tau_\chi\bar\xi^4+4\tau_\gamma
\bar\sigma^2\bar\xi^2\ear
with $m=m_u+m_d+m_s$. The sign of $\zeta$ may be positive or
negative\footnote{Only the relative phase between $\zeta$ and $m$
is relevant. Conservation of $C,P$ requires that the phases of
$\zeta$ and $m$ are equal up to a minus sign and we take both
$\zeta$ and $m$ real. We choose a phase convention such
that the expectation value of $\bar\sigma$ is positive and
concentrate on the case $m_q>0$.} and we discuss below acceptable
scenarios for both cases.
For positive couplings $\lambda_l,\tau_l$
the classical potential typically\footnote{This clearly
holds if $\zeta\bar\sigma$ is positive at the minimum.
For negative $\zeta\bar\sigma$ one may also envisage the possibility
that already the effective action (\ref{2}) has its minimum
for $\bar \xi\not=0$.}
has its minimum for $\bar\xi=0$. Then spontaneous symmetry
breaking of color symmetry can only be induced by the low momentum
fluctuations.

The fermionic functional integral is easily evaluated and gives
a contribution to the effective scalar potential
\be\label{18}
U(\bar\sigma,\bar\xi)=U_{cl}(\bar\sigma,\bar\xi)+
\Delta_q U(\bar\sigma,\bar\xi)\ee
For a sharp ultraviolet cutoff $k$ this reads
\be\label{19}
\Delta_q U=-\frac{1}{8\pi^2}\int^{k^2}_0dxx[8\ln(x+M^2_8)+\ln
(x+M^2_1)]\ee
and one observes that $\Delta_q U$ respects the (accidental?)
symmetry\footnote{This is not trivial since an invariant
$\epsilon_{a_1a_2a_3}\epsilon_{b_1b_2b_3}
\epsilon_{i_1i_2i_3}\epsilon_{j_1j_2j_3}
\chi_{i_1j_1,a_1b_1}\chi_{i_2j_2,a_2b_2}\\ \chi_{i_3j_3,a_3b_3}
+c.c.$ is consistent with color and chiral symmetries as well as $P$
and $C$.}
of the classical potential $\bar\xi\to-\bar\xi$.
The quark fluctuations tend to
destabilize the minimum of $U_{cl}$ at $\bar\sigma=\bar\xi=0$
for $m=0$, since nonzero values of $M_8,M_1$ are preferred.
They give a negative contribution to the quadratic term
obtained by an expansion of $\Delta_qU$ for small $\bar\sigma$
and $\bar\xi$
\be\label{FF}
\Delta_qU=-\frac{k^2}{2\pi^2}(9\lambda^2_\sigma\bar\sigma^2+
\frac{4}{3}\lambda_\chi^2\bar\xi^2)+...\ee
For $\lambda_\sigma k^2>4\pi^2/3\ ,\ \lambda_\chi k^2>4\pi^2$ they
overwhelm the classical mass terms $6\lambda_\sigma\bar\sigma^2$,
$(8/3)\lambda_\chi\bar\xi^2$ for $\bar\sigma$ and $\bar\xi$,
respectively. We conclude that the
quark fluctuations are the main driving force
for spontaneous chiral symmetry breaking. Furthermore, the cubic
``anomaly terms'' favor the condensates even in presence of small
enough positive quadratic terms at the origin.
In mean field theory one has to determine the
minimum of $U$. We have done this by numerically solving
the field equations. In this context we note the
simple closed form of the contribution from $\Delta_q U$, namely
\bear\label{20}
&&\frac{\partial}{\partial\bar\sigma}\Delta_q U=8A_8\frac{\partial M_8}
{\partial\bar\sigma}+A_1\frac{\partial M_1}{\partial \bar\sigma}
\quad,\quad
\frac{\partial}{\partial\bar\xi}\Delta_q U=8A_8\frac{\partial M_8}
{\partial\bar\xi}+A_1\frac{\partial M_1}{\partial \bar\xi}
\nonumber\\
&&A_{1,8}=\frac{M^3_{1,8}}{4\pi^2}\left\{\ln
(1+\frac{k^2}{M_{1,8}^2})-\frac{k^2}{M^2_{1,8}}\right\}\ear
For a large range of couplings we find indeed a nonzero expectation
value for the octet condensate $\bar\xi$!

As long as the effective couplings $\lambda_l,\tau_l,\zeta$ are not
computed from QCD, the predictive power of the present approach
is limited. We will be satisfied here by presenting a set of perhaps
reasonable values for these couplings for which a realistic
phenomenology can be obtained. Some of the parameters are tuned
to fit with observed masses. We choose $k=850$ MeV equal to the average
of the $\rho$-meson or gluon mass. The renormalization scale for the
quark mass is taken at $\mu=1$ GeV in the vicinity of the baryon
masses, and we take for the quark mass sum $m(\mu)=220$ MeV.
Our selected parameters are
\bear\label{20a}
&&\lambda_\sigma k^2=53(63,57),\quad
\lambda_\chi k^2=0.47(0.9,5.8),\quad \zeta k^5=356(32,-73.5),\\
&&\tau_\sigma k^8=4700(185,8060),\quad \tau_\chi k^8=87(44,314),
\quad \tau_\gamma k^8=5200(350,1570)\nonumber\ear
Here the numbers in
brackets refer to two sets with  inclusion of fluctuations of
the gauge bosons and the pseudoscalar octet, as discussed
below. For these values the minimum of $U$ occurs for
\be\label{21}
<\bar\sigma>=\frac{1}{2}(235\ {\rm MeV})^3\quad,\quad
<\bar\xi>=\frac{1}{2}(400(470,280)\ {\rm MeV})^3\ee
and we observe the direct relation of $<\bar\sigma>$ with the (singlet)
quark-antiquark condensate\footnote{In our Euclidean conventions
$<\bar\psi\psi>$ corresponds to the expectation value of
$\bar\psi\bar\gamma\psi$.}
and the parameters $B$ and $f$ of chiral
perturbation theory ($f$ is the meson decay constant)
\be\label{22}
<\bar\sigma>=-\frac{1}{2}<\bar\psi\psi>(\mu)=\frac{1}{2}
B(\mu)f^2\ee
The average mass of the lightest baryon octet and singlet obtains by
evaluating eqs. (\ref{15}), (\ref{16}) for the
expectation value (\ref{21}). For our parameters they coincide
with the observed value of the average octet mass and an
arbitrary fixed singlet mass.\footnote{The baryon singlet
is presumably very broad and it is not obvious if it should
be associated with an observed resonance. The lowest mass
resonance with the correct quantum numbers occurs at 1.6 GeV.}
\be\label{23}
M_8=1.15\ {\rm GeV}\quad,\quad M_1=1.4\ {\rm GeV}\ee

It is instructive to determine the masses of the most important
bosonic excitations for our parameter set. For
a computation of the scalar masses one needs the scalar wave
function renormalizations which have not been computed so far.
Altogether, the fields $\sigma$ and $\xi$ contain 162 real scalars
which transform under $SU(3)$, $P$ and $C$ as
\bear\label{24}
&&2\times(1^{-+}+1^{++}+8^{-+}+8^{++}+8^{--}+8^{+-})\nonumber\\
&& +10^{--}+10^{+-}
+\bar{10}^{--}+\bar{10}^{+-}+27^{-+}+27^{++}\ear
One octet is absorbed by the Higgs mechanism into the longitudinal
component of the massive gluons. The representations $10, \bar{10}, 27$
contain scalars with electric charge two. Many of these states
may be too broad to be detected experimentally
as resonances. Of particular interest are
the lightest pseudoscalar octet $8^{-+}$ which  corresponds
to massless Goldstone bosons for $m_q\to0$, and the lightest
pseudoscalar singlet $1^{-+}$ associated with the $\eta'$-meson.
They are most easily described in a nonlinear
representation \cite{RU}
\bear\label{25}
\sigma_{ab}&=&<\bar\sigma>U_{ab}\quad,\quad\xi_{ij,ab}=\frac{1}{
\sqrt6}<\bar\xi>(U^{1/2}_{ai}U_{jb}^{1/2}-\frac{1}{3}\delta_{ij}U_{ab}),
\nonumber\\
U&=&\exp(-\frac{i}{3}\theta)\exp(i\frac{\Pi^z\lambda_z}{f})\quad,
\quad U^\dagger U=1\ear
If we parametrize the effective scalar kinetic
terms\footnote{Presumably the dominant contribution
to the kinetic terms is induced by quantum fluctuations with
$q^2<k^2$. It can also be computed in the mean field
approximation. Integrating out first the fermions will
also induce an effective gauge coupling between the gluons
and the color octet $\xi$, replacing $\partial_\mu\xi$ by
a suitable covariant derivative.}by
\be\label{26}
{\cal L}_{s,kin}=Z_\sigma\partial^\mu\sigma_{ab}^*\partial_\mu
\sigma_{ab}+Z_\xi\partial^\mu\xi^*_{ij,ab}\partial_\mu\xi_{ij,ab}\ee
the kinetic term for the pseudoscalar nonet reads
\bear\label{27}
{\cal L}_{U,kin}&=&(Z_\sigma<\bar\sigma>^2+\frac{7}{36}Z_\xi<\bar\xi>^2)
\quad Tr\partial^\mu U^\dagger\partial_\mu U\nonumber\\
&=&\frac{1}{4}f^2\ Tr \partial^\mu U^\dagger\partial_\mu U\ear
An average meson decay constant $f=106$ MeV
corresponds to
\bear\label{27a}
&&Z_\sigma+\frac{7}{36}Z_\xi(\bar\xi/\bar\sigma)^2
=Z_\sigma(1+x)=(350\ {\rm MeV})^{-4},\nonumber\\
&&
x=\frac{7}{36}\frac{<\bar\xi>^2}{<\bar\sigma>^2}\frac{Z_\xi}{Z_\sigma}
\ear

If this is the case, chiral symmetry
guarantees realistic masses for the pions, kaons, and the $\eta$-meson.
Neglecting $SU(3)$-violating effects for the
expectation values they are given by
\bear\label{27b}
M^2_\pi&=&2<\bar\sigma>(m_u+m_d)/f^2,\nonumber\\
M^2_K&=&2<\bar\sigma>(m_u+m_s)/f^2\ear
independently of details of the potential \cite{QM}. For
realistic $f$ and appropriate quark mass ratios
$m_u/m_s, m_d/m_s$ the expectation values (\ref{21})
lead to the observed light pseudoscalar meson masses.

If we associate the $\eta'$-meson with $\theta$ in eq. (\ref{25}), neglect
its mixing with other mesons with the same quantum numbers and
assume that its kinetic term is dominated by eq. (\ref{27}) we
find
\be\label{28}
M^2_{\eta'}=\frac{36\zeta}{f^2}
(\bar\sigma^3-\frac{2}{9}\bar\sigma\bar\xi^2+\Delta_{\eta'})+m^2_g\ee
where $m_g=410$ MeV is the contribution from explicit chiral symmetry
breaking due to the current quark masses and
\bear\label{29}
\Delta_{\eta'}&=&8A_8(\bar\sigma^2-\frac{2}{27} \bar\xi^2+\frac{1}
{9\sqrt6}\bar\sigma\bar\xi)\cdot
(2\lambda_\sigma\bar\sigma-\frac{2}{3\sqrt6}\lambda_\chi\bar
\xi+6\tau_\sigma\bar\sigma^3\nonumber\\
&&+\frac{4}{3}\tau_\gamma\bar\xi^2\bar\sigma-\frac{1}
{\sqrt6}\tau_\gamma\bar\sigma^2\bar\xi-\frac{8}{9\sqrt6}\tau
_\chi\bar\xi^3)
+A_1(\bar\sigma^2-\frac{2}{27}\bar\xi^2-\frac{8}{9\sqrt6}
\bar\sigma\bar\xi).\nonumber\\
&&(2\lambda_\sigma\bar\sigma+\frac{16}{3\sqrt6}
\lambda_\chi\bar\xi
+6\tau_\sigma\bar\sigma^3+\frac{4}{3}\tau_\gamma\bar\xi^2\bar\sigma+
\frac{8}{\sqrt6}\tau_\gamma\bar\sigma^2\bar\xi+\frac{64}
{9\sqrt6}\bar\xi^3)\ear
arises from the $\theta$-dependence of the fermionic fluctuation
determinant giving rise to $\Delta_q U$. As is should be, $M_{\eta'}$
vanishes for zero quark masses and vanishing chiral anomaly
$(\zeta=0)$. For our parameters eq. (\ref{28}) yields the observed value
$M_{\eta'}=960$ MeV.

The gluons are coupled to the colored scalar fields $\xi$ and
therefore acquire a mass through the Higgs mechanism. We identify
this mass with the average mass of the light vector meson
octet $M_\rho=$850 MeV. Denoting the effective gauge coupling by $g$ one
obtains
\be\label{30}
M_\rho=g\ Z^{1/2}_\xi<\bar\xi>=\frac{3}{\sqrt7}\left(\frac{x}{1+x}\right)
^{1/2}gf\ee
For our set of parameters one finds a large value of $x$ if
$Z_\sigma/Z_\xi$ is not too large
For $Z_\sigma/Z_\xi=0.44(3,0.13)$ this yields\footnote{The relation to the
parameters used in \cite{GM} is given by $m^2_\phi=2\lambda_\sigma
/Z_\sigma=(5.1(3.6,3.56)\ {\rm GeV})^2,\ m^2_\chi=2\lambda_\chi/Z_\xi=(
0.32(0.75,0.42)\ {\rm GeV})^2,\ h=2\lambda_\sigma Z_\sigma^{-1/2}
=62(48,46),\ \tilde h=2\lambda_\chi Z_\xi^{-1/2}=0.36(1.2,1.8),\
\tilde\epsilon_\phi=m^2_\phi/(h^2k^2)=(2\lambda_\sigma k^2)^{-1}
=0.009(0.008,0.009),\tilde\epsilon_\chi=
m^2_\chi/(\tilde h^2k^2)=(2\lambda_\chi k^2)^{-1}=1(0.55,0.086)$, with  
$\varphi=Z_\sigma^{1/2}
\sigma,\ \chi=Z^{1/2}_\xi\xi$.}
\be\label{32}
x=10.9(4,4),\ g=7.4(7.9,7.9),\ Z_\xi=(530(690,320)\ {\rm MeV})^{-4}\ee

One may want to include the effects of gauge boson and scalar
fluctuations in the effective potential. In the mean field
approximation one finds
\bear\label{34}
\Delta_gU&=&\frac{3}{4\pi^2}\int^{k^2}_0 dx\ x\ ln(x+g^2Z_\xi\bar\xi^2)
\nonumber\\
\Delta_sU&=&\frac{1}{32\pi^2}\sum_s\int^{k^2}_0 dx\ x\ ln
(x+M^2_s)\ear
for the gauge bosons and scalars, respectively. Here $s$
counts the scalars or pseudocscalars with mass $M_s$
depending on $\bar\sigma$ and $\bar\xi$. Additional uncertainties
arise from the possible $(\bar\sigma,\bar\xi)$-dependence of
the wave function renormalizations $Z_{\sigma,\xi}$ and $g$.
We take here $Z_\sigma$ and $Z_\xi$ constant and use
$\partial g/\partial\bar\xi=\eta_gg,\partial g/\partial\bar\sigma=0$.
As long as $\bar\xi^A,\ A>0$, acts as an infrared cutoff for the
gauge boson fluctuations, the quantity $\eta_gg$ is related to the
non-perturbative  $\beta$-function of $g$. From asymptotic freedom
one expects $\eta_g$ to be negative. Typically $|\eta_g(g)|$ increases
from small perturbative values $\sim g^2$ for small $g$ to large
values in the range of large $g$. The effective $\bar\xi$-dependent
gluon mass obeys $\partial M_\rho(\bar\xi)/\partial\bar\xi=(1+\eta_g)
M_\rho(\bar\xi)$ and therefore has a minimum for $\eta_g=-1$.
One concludes that $\Delta_gU$ is probably minimal for $\bar\xi>0$,
namely for $\eta_g(g(\bar\xi))=-1$. For $\eta_g(\bar\xi\to0)
<-1$ also the gauge boson fluctuations
favor color symmetry breaking. For our numerical solution we show
the results for $\eta_g(<\bar\xi>)=-1.9, -1.15$.
(We mention that for $\eta_g(<\bar\xi>)=-1$ the contribution from
$\Delta_gU$ drops out in the solution of the field equations.)

For the scalar fluctuations we have only included the light pseudoscalar
octet. According to (\ref{27b}) and (\ref{27}) one finds $M^2_s
\sim\bar\sigma/(\bar\sigma^2+\frac{7}{36}
\frac{Z_\xi}{Z_\sigma}\bar\xi^2)$. For given $\bar\sigma$ and
$Z_\xi/Z_\sigma$ the pseudo-Goldstone boson fluctuations $\Delta_sU$ favor
again nonzero $<\bar\xi>$ since $M_s(\bar\xi\to\infty)\to0$.
Two parameter sets for a numerical solution including the effects
of the gauge bosons and pseudo-Goldstone bosons are given in
brackets in eq. (\ref{20a}), one with positive
and one with negative $\zeta$. (For the last set
we use $k=820$ MeV.) We also have shown above in brackets
the corresponding values for various quantities -- the absence of
brackets indicating that parameters are tuned in order to obtain
physical values.

The choice of the parameters in $\tilde U_{k}$ for which
we have presented
results is somewhat arbitrary -- we also have found
acceptable solutions
for rather different sets. The phenomenologically acceptable
couplings seem not unnatural and we conclude that spontaneous color
symmetry breaking is a plausible phenomenon in QCD. Two features
seem characteristic for multiquark interactions that lead to
phenomenologically acceptable solutions with the observed values
of $M_8, M_1,M_\pi, M_K,M_\eta,M_{\eta'}$ and $M_\rho$. The
first is a large ratio $\lambda_\sigma/\lambda_\chi$. This may
be explained by the different running of the effective couplings
$\lambda_\sigma$ and $\lambda_\chi$.
The second is the large value of the effective gauge coupling $g$
between the gluons and color octet bilinear $\xi\sim\tilde\chi$.
At first sight this seems somewhat surprising in view
of the fact that $M_\rho\approx850$ MeV is expected to act as
an infrared cutoff. Extrapolations of the two- or three-loop
$\beta$-functions in the $\overline{MS}$-scheme into this
domain would yield substantially smaller values of $\alpha_s(M_\rho)$.
An understanding of this issue needs, however, information about
the relation of our effective coupling $g$ to the
perturbative gauge coupling
in the $\overline{MS}$ scheme, and about the running of $g$ in
the non-perturbative domain. A first discussion of this issue
can be found in \cite{RU}. We find it reassuring
that the gauge boson fluctuations seem not to play a dominant
role for the dynamics of spontaneous color symmetry breaking.

In conclusion, we have presented a consistent mean field picture for
gluon-meson duality. The quantitative details are not yet
all settled. They would need a computation of the multiquark
interactions $\tilde U_k$. Without such a computation in the
framework of QCD no definite conclusion on the issue of
``spontaneous color symmetry breaking'' can be drawn. Nevertheless,
it is remarkable that for reasonable multiquark couplings
a simple scenario can successfully describe all masses of the
light pseudoscalar and vector mesons and the baryons, as well as
their interactions.

We have identified three mechanisms by which
fluctuations operate in favor of
a nonvanishing expectation value $<\bar\xi>$ for the color
octet quark-antiquark condensate (in a fixed gauge). Perhaps the most
important one arises from the fermion fluctuations which favor
large baryon masses and therefore nonzero $\bar\sigma,\bar\xi$
due to the minus sign in eq. (\ref{19}). They induce a negative
contribution to the quadratic term $\sim\bar\xi^2$ in the effective
potential. The second ingredient is due to the fluctuations of the
(pseudo-)Goldstone bosons. Now $\Delta_sU$ favors small meson masses
or a large decay constant $f$, and therefore again large $\bar\sigma,
\bar\xi$. Finally, a third mechanism is possible if the gluon mass is
minimal for a nonzero value of $<\bar\xi>$. This last issue depends
on unknown properties of the non-perturbative running of the gauge
coupling and is therefore less solid than the first two mechanisms.
The combined effect of these mechanisms has to compete with the
``classical mass terms'' $\sim \lambda_\sigma,
\lambda_\chi$ which favor the absence of spontaneous chiral and
color symmetry breaking. The crucial dynamical question
is whether the fluctuation effects are strong enough.
Finally, we also mention a fourth mechanism which becomes
possible for $\zeta<0$. An increase in $<\bar\sigma>$
due to fluctuations lowers the mass term for $\bar\xi$
if $\zeta<-9\tau_\gamma\bar\sigma$ (\ref{17}). We find it
plausible that spontaneous color symmetry breaking indeed
occurs in QCD.

The quantitative validity of mean field theory remains questionable
in the strongly non-perturbative domain discussed in this note. Mean
field theory should, however, indicate at least the qualitative
tendencies of the fluctuation effects. We have included here the
most important fluctuations of the light baryons and mesons. Furthermore,
we find that the term $2\lambda_\sigma\bar\sigma$ accounts for 82(99,95)
precent of the octet mass $M_8$. This suggests that the neglected
effects of the residual multi-quark interactions after bosonization
are subleading. Finally, the existence of a local minimum with
$<\bar\sigma>\not=0,\ <\bar\xi>\not=0$,
as found in this note, only requires the validity of mean field
theory in a neighborhood around this minimum. Statements of this
type should be more robust than a mean field computation of
global properties of the effective potential for all $\bar\sigma,
\bar\xi$. We recall that the main shortcoming of mean field theory
is that it neglects the effective running of the couplings, as
fluctuations with different momenta are included. In the vicinity
of the minimum at $(<\bar\sigma>,<\bar\xi>)$ the only
excitations with mass much smaller than $k$ are the pions.

It would be interesting if some qualitative results of this
investigation, namely that the color octet condensate
$<\bar\xi>$ dominates
$M_\rho,f$ and $M_1-M_8$, whereas the singlet condensate $<\bar\sigma>$
essentially determines $M_8$ and the combination
$M_s^2f^2$ for the pseudoscalars, could be
seen in future lattice simulations. This may be possible
indirectly by a study of the quark mass dependence of the meson
and baryon masses.

\end{document}